# Coherently combined 16 channel multicore fiber laser system

A. KLENKE,[1,2,*] M. MÜLLER,[1] H. STARK,[1] F. STUTZKI,[3] C. HUPEL,[3] T. SCHREIBER,[3] A. TÜNNERMANN[1,2,3] AND J. LIMPERT[1,2,3]

[1]*Institute of Applied Physics, Abbe Center of Photonics, Friedrich-Schiller-Universität Jena, Albert-Einstein-Straße 15, 07745 Jena, Germany*
[2]*Helmholtz-Institute Jena, Fröbelstieg 3, 07743 Jena, Germany*
[3]*Fraunhofer Institute for Applied Optics and Precision Engineering, Albert-Einstein-Str. 7, 07745 Jena, Germany*
*\*Corresponding author: a.klenke@gsi.de*

**We present a coherently combined laser amplifier with 16 channels from a multicore fiber in a proof-of-principle demonstration. Filled aperture beam splitting and combination together with temporal phasing is realized in a compact and low-component-count setup. Combined average power of up to 70 W with 40 ps pulses are achieved with combination efficiencies around 80%.** © 2018 Optical Society of America under Open Access Publishing Agreement. Users may use, reuse, and build upon the article, or use the article for text or data mining, so long as such uses are for non-commercial purposes and appropriate attribution is maintained. All other rights are reserved.

The utilization of coherent combination has drastically increased the performance figures like average power, pulse energy, and peak power of fiber laser systems. This could be realized by distributing power scaling challenges across multiple emitters to overcome the different physical limitations of these values. For example, limitations include average power related effects such as mode-instabilities [1–3], but also non-linear effects for parameters such as pulse energy and peak power. Today, single amplifier channels have been outperformed by more than one order of magnitude by systems employing either spatial combination of multiple amplifiers [4], temporal combination of multiple pulse replicas [5] or even both [6]. Notable results include a system emitting 12 mJ femtosecond pulses at 700 W average power realized with 8 parallel channels and 4 temporal pulse replicas [7] or a 4 kW continuous-wave system with 8 channels [8]. According to theoretical calculations, systems using even greater parallelization appear to be viable, since the expected total combination efficiency converges with increasing number of channels N [9,10]. This offers the potential to move fiber laser systems towards performance parameters (e.g. J-class pulse energies at high repetition rates) required for demanding applications in high-field physics, such as laser particle acceleration.

However, in most of the previously demonstrated laser systems based on coherent combination, the component count, and thus the complexity of the system basically grows linearly with the number of channels for spatial combination. In alleviated terms, this statement also holds true for temporal pulse combination. Therefore, it is necessary to employ integrated multi-channel components in order to decouple the component count from the channel count. The integration of multiple amplification channels into a multicore fiber has already been investigated for tiled-aperture combination in a low-power experiment [11]. In this case, a spatial light modulator is used to generate a beam pattern matching the core positions in the multicore fiber and also to set the phase of the specific beams. 860 fs femtosecond pulses at 100 kHz repetition rate were achieved. However, due to the limited fill factor of the tiled-aperture combination scheme, the power in the central feature is theoretically limited to 76% and a value of 49% was realized experimentally. Using diffractive-optical-elements (DOEs), the combination efficiency can be increased by implementing the filled-aperture technique [12]. However, the diffractive nature of the combination element makes this approach not suitable for femtosecond sources without a pre-compensation of the spatial chirp for each incident beam. Another method is to use multiple coupled cores [13], but here the power scalability especially regarding the mode-instability threshold and the possibility to use them as amplifiers remains unclear. Additionally, the spectral beam combination approach has been demonstrated with multicore fibers [14,15].

In this paper, we present the realization of a filled-aperture combination setup with a 16 core fiber. The filled-aperture scheme allows for high combination efficiencies and a near-diffraction-limited output beam quality. Additionally, the non-diffractive nature of the employed components allows for application to broadband sources, i.e. ultrashort-pulse operation, in the future.

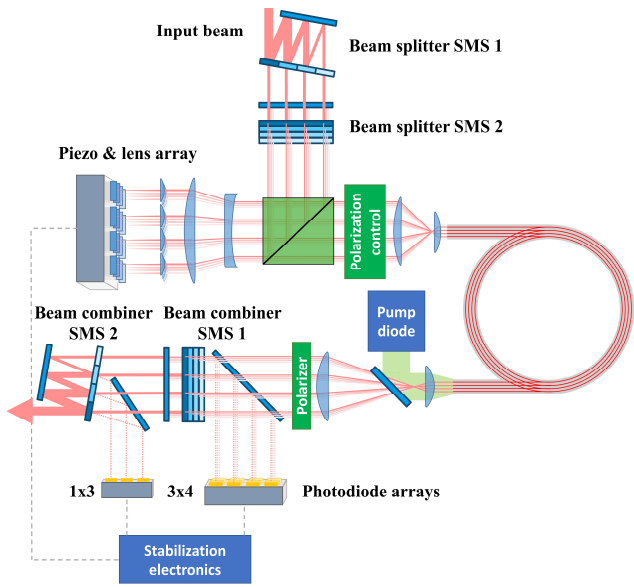

Fig. 1. Schematic setup of the amplifier, including the two beam splitters SMS 1 and SMS 2, piezo array for beam phasing, the 16 core fiber, beam combiners SMS 1 and SMS 2 and phase detection system

The setup, shown schematically in Fig. 1, consists of multiple components that feature a multi-channel design. A laser with 40 ps pulse duration, 20 MHz repetition rate and 300 mW average power operating at a wavelength of 1032 nm is used as a seed system. The emitted beam first passes through a segmented-mirror-splitter (SMS). This component consists of one high-reflective mirror and one element with zones of different reflectivity. For this 16 channel setup, there are 4 zones with reflectivities of 0% (AR-coated), 50%, 66% and 75%. By placing both elements parallel to each other the input beam is split into 4 parallel equidistant beams (Beam splitter SMS 1 in Fig. 1). A transmissive beam splitter based on a similar design was presented in [16]. This splitting process is repeated again with a second SMS (Beam splitter SMS 2) rotated by 90°. Thus, each of the 4 beams is split into 4 beams again, resulting in a two-dimensional output beam array with 4 mm pitch between the beams. The beams are reflected at a polarization beam splitter cube (PBS) towards a 4x4 piezo array with mirrors attached to each piezo actuator. Due to required pitch of 9 mm of the piezo array, the beam array is magnified by a telescope. Each piezo has a piston range of 20 µm, resulting in an effective value of 40 µm in the presented double-pass configuration. While great care has been taken when attaching the mirrors to the piezos, a small static tilt is still present in addition to a dynamic tilt that occurs during actuator movement. This would result in a distortion of the beam array. Therefore, a lens array with 100 mm focal length is placed in front of the piezo array, placing the piezo mounted mirrors into the Fourier domain of the beams. This configuration transfers the tilt into a transversal beam offset, which is small compared to the beam sizes, hence alleviating the beam array distortion. Additionally, spherical aberrations of the telescope are reduced due to the horizontal and vertical flipping of each beam in the double-pass configuration. The beam array is directly imaged to the end-facet of the employed multicore fiber with an optical ensemble in 4f-configuration after passing through the PBS.

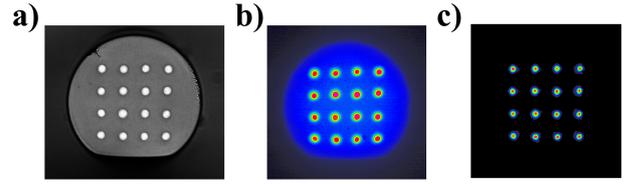

Fig. 2. Image of the cleaved end-facet of the multicore fiber (a), corresponding amplified-spontaneous emission (ASE) output (b), and amplified signal output (c)

The in-house designed fiber consists of 16 step-index Ytterbium-doped cores in a 4x4 configuration shown in Fig. 2. Each core has a diameter of 19 µm with a pitch of 55 µm between the core centers and a numerical aperture of 0.06. Based on previous research on a 4 core fiber regarding the average power handling [17], this ratio between core size and pitch is sufficient to suppress optical and thermal coupling between the cores, and, therefore, avoid a detrimental impact on the mode-instability threshold. Thus, the multicore fiber can be treated like multiple spatially separated fibers. As can be seen in Fig. 2, the core positions in the manufactured fiber show small deviations in the range of a few µm from the perfectly rectangular pattern. The consequences of this issue will be discussed later in this manuscript. The multicore-fiber has an outer diameter of 340 µm and a low-index polymer coating acting as the shared pump cladding. Additionally, the fiber was given a D-shape in order to guarantee mode mixing of the pump light coming from a 200 µm fiber coupled 976 nm pump diode. A fiber length of 5 m was chosen for sufficient pump absorption (>97% for 190 W pump power) and the end-facets were angle-polished to avoid back-reflections. After amplification, the output beam array is again magnified with optics in 4f-configuration to the initial pitch of 4 mm (corresponding to beam diameters of 1.4 mm) before passing through another two SMS elements for beam combination. The distance between the two elements of each SMS at this combination stage is matched to the splitting stage, to provide equal path lengths for each beam. At the first SMS (Beam combiner SMS 1 in Fig. 1), the 16 incoming beams are combined into 4 output beams by combining 4 columns of 4 beams each in vertical direction. In each of these 4 columns this results in 3 beam interference steps and thus, in 3 beams containing the parts that do not interfere constructively. Therefore, the first rejection port is formed by a 3x4 beam array, which represents the losses of the beam combination process at this stage. A small fraction of this array is guided towards a 3x4 photodiode array connected to the active phase stabilization system, while the bulk of the power is dumped. At the second SMS (Beam combiner SMS 2 in Fig. 1), the 4 remaining beams are further combined into a single output beam in horizontal direction, thus resulting in a 1x3 beam array emitted from the second rejection port. Again, a fraction of the rejected light is guided towards a 1x3 photodiode array. It should be noted that during the propagation through the SMS, the high-power beam is always reflected, which increases the average power handling capabilities. Actually, this setup is very similar to cavity-enhancement setups, where the handling of MW-level average powers has already been demonstrated [18].

The required active path-length stabilization is realized by applying small sinusoidal phase modulations to the piezo array. The top-left actuator remains steady, allowing to use the corresponding beam as the reference signal. A modulation frequency of 6 kHz is applied to the 3 other actuators in the first row, while a frequency of 4 kHz is chosen for the 12 actuators in the remaining 3 rows. Additionally, a phase offset of π is added between the phase modulations of adjacent actuators with the same frequency, which reduces residual phase fluctuations in the combined beam. The demodulated error signal was evaluated with a bandwidth of 1 kHz. This multi-detector scheme allows to stabilize the 16 beams of the setup even with a limited total bandwidth of the piezo actuators of around 10 kHz by reusing the same two modulation frequencies for multiple actuators [19].

The fiber does not contain any structures to preserve the polarization state and, thus, polarization changes for different cores were observed even in a straight piece of fiber. In the experimental setup, the fiber was bent to a radius of about 30 cm. Therefore, static polarization control was necessary for the cores realized by an array of quartz waveplates. This allowed to have 85% of power in p-polarization after amplification for all average powers.

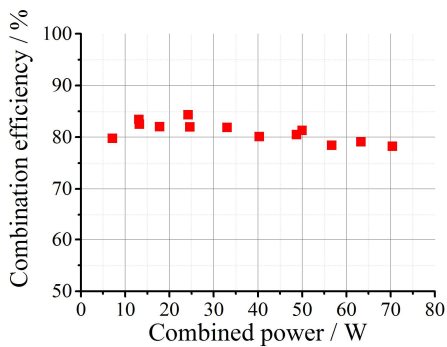

Fig. 3. Combination efficiency depending on the combined average power

The most prominent way to determine the quality of the coherent combination process is to measure the combination efficiency. In this case – to distinguish between depolarization in the amplifying fiber and the performance at the SMS beam combiner - it is defined as the output power in the combined output over the total amplified power after the polarizer. In Fig. 3, the combination efficiency is shown depending on the combined power of up to 70 W, limited by the power of the employed pump diode. As can be seen, the results are in a range of a few percentage points around 80%, with a small reduction for higher average powers. Both aspects, the upper limit of the efficiency as well as this reduction are investigated in the following.

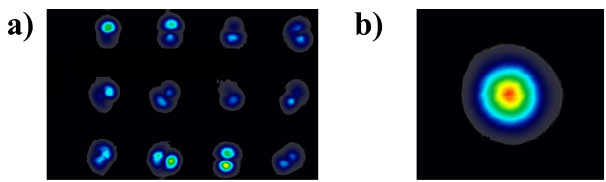

Fig. 4. Image of the beams at the first rejection port when stabilized (a), Beam profile of the combined beam (b)

Considering the overall efficiency, losses during the combination process can be observed by investigating the non-combining parts of the beams. Fortunately, this is easily realizable in the presented setup by directing a fraction of the power from the two rejection ports to cameras (same beams as for the photodiode arrays). In Fig. 4 (a), the image from the first rejection port is shown with high exposure time. The non-combining parts have mostly higher-order mode shape, which can be explained by the presence of varying higher-order mode content in the different signal cores and residual beam alignment errors. The stabilization system optimizes for maximum combined power, meaning the system is optimized for constructive interference of the dominant fundamental mode, while the amplitude and phase of the higher-order modes is random. Therefore, higher-order mode content is suppressed in the combined beams and appears as losses at the rejection port. Supporting this explanation, changing the coupling of the beams changes the observed pattern. The previously mentioned core position deviations in the fiber result in an increasing coupling into higher order modes of the cores instead of the fundamental mode, and, therefore, increase these losses. Hence, improved production processes with better core positioning have the potential to improve the combination efficiency.

The small reduction of efficiency at higher average powers can be explained by measuring the optical spectrum of the output. In this experiment, spectral broadening caused by self-phase modulation from 130 pm to 260 pm at the highest average power was observed. Therefore, differences between the input and output powers in each core result in core dependent spectral width and phases, which reduce the combination efficiency [10]. This effect was observed by an increased sensitivity at higher powers regarding the beam coupling into the multicore fiber.

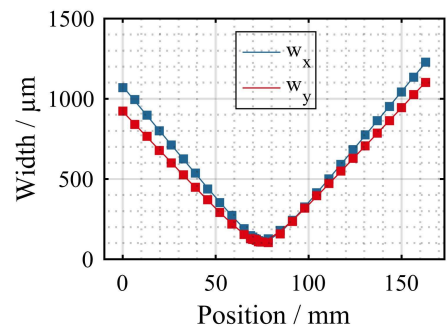

Fig. 5. Caustic of the combined beam for the $M^2$ measurement

In order to specify the quality of the combined beam (Fig. 4 (b)), which is an advantage of the filled-aperture scheme, an $M^2$ measurement device was employed. At full power, a near diffraction-limited value of $M^2 < 1.2 \times 1.1$ was measured for the beam caustic shown in Fig. 5. This value also compares favorably to the observed beam quality of single channels where values of up to 1.4 were measured, supporting the described beam-cleaning effect.

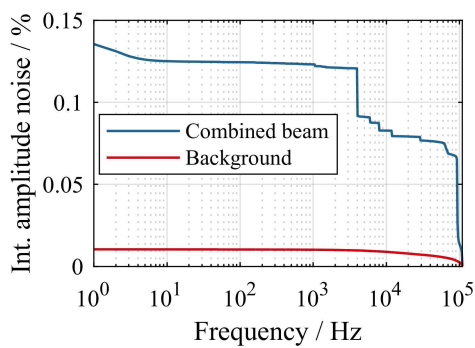

Fig. 6. Integrated amplitude noise of the combined beam and of the measurement background for a frequency range of 1 Hz to 100 kHz

The power stability of the combined output was determined by connecting a photodiode through a 240 kHz low-pass filter (to suppress the fundamental repetition rate) to a high resolution oscilloscope and measuring the output signal with a sampling rate of 1 MSp/s. In Fig. 6 the integrated amplitude noise is shown for the combined beam and for the background noise for a frequency range of 1 Hz to 100 kHz. The step-like increase of the integrated noise at 4 kHz, 6 kHz and their harmonics are due to the applied phase modulations. A relative intensity noise (RIN) of 0.14% was calculated for the combined signal and a value of 0.01% for the background noise. This very small integrated noise demonstrates that the stabilization successfully compensates for phase perturbations of the interferometer channels.

In conclusion, we have demonstrated a scheme for the implementation of a filled-aperture coherent combination with a multicore fiber. Up to 70 W of pump limited average power with near-diffraction limited beam quality could be achieved at combination efficiencies of around 80%. The combination is realized by integrated multi-channel elements implementing beam splitting, combination, and phasing. Hence, an increase of the number of channels can be achieved without increasing the number of discrete optical components. In future research, the combination of femtosecond pulses in such a setup has to be investigated. This requires a matching of the path-lengths in the different channels to a few wavelengths in order to guarantee the temporal overlap of such short pulses. Additionally, the average power handling regarding the mode-instability threshold will have to be further investigated for the presented two-dimensional core layout. Finally, techniques to manufacture large-area-mode multicore fibers [17] (e.g. micro-structured fibers) have to be developed and the number of cores in the fiber has to be increased. The presented concept can then scale the performance of high repetition-rate femtosecond fiber based laser systems to a new level, such as moving from the current mJ-level pulse energies to the J-level, which will enable a variety of new applications.

**Funding.** European Research Council (ERC) (670557), "MIMAS"; Free State of Thuringia (2015FE9158), "PARALLAS"; Fraunhofer research cluster "Advanced Photon Sources".

**Acknowledgment**. We thank the IPHT Jena for drawing our preform to fibers. M.M. acknowledges financial support by the Carl-Zeiss-Stiftung.